\documentstyle[aaspp4]{article}
\input{epsf}

\newif\ifAMStwofonts
\AMStwofontstrue

\def\be{\begin{equation}}
\def\ee{\end{equation}}
\def\ba{\begin{eqnarray}}
\def\ea{\end{eqnarray}}
\def\go{\mathrel{\raise.3ex\hbox{$>$}\mkern-14mu
             \lower0.6ex\hbox{$\sim$}}}
\def\lo{\mathrel{\raise.3ex\hbox{$<$}\mkern-14mu
             \lower0.6ex\hbox{$\sim$}}}
\def\Oms{\hat{\Omega}_{\rm s}}
\def\Omsb{\Omega_{\rm s}}
\def\Omsi{\hat{\Omega}_{\rm s0}}
\def\Omsib{\Omega_{\rm s0}}
\def\oma{\sigma_{\rm A}}
\def\nus{\nu_{\rm s}}
\def\nusi{\nu_{\rm s0}}
\def\omi{\omega}
\def\omr{\sigma}
\def\nui{\nu}
\def\nuii{\nu_{\rm 0}}
\def\alphai{\alpha_{\rm 0}}
\def\lp{\left(}
\def\rp{\right)}
\def\lb{\left[}
\def\rb{\right]}
\def\tI{\tilde{I}}
\def\tJ{\tilde{J}}
\def\th{\tilde{h}}
\def\Jc{J_{\rm c}}
\def\taugr{\tau_{\rm GR}}
\def\tauv{\tau_{\rm V}}
\def\taum{\tau_{\rm M}}
\def\tauem{\tau_{\rm EM}}
\def\taua{\tau_{\rm A}}
\def\va{v_{\rm A}}
\def\Alfven{Alfv\'{e}n }
\def\tsat{t_{\rm sat}}
\def\tsato{t_{\rm sat0}}
\def\xsat{x_{\rm sat}}
\def\ysat{y_{\rm sat}}
\def\Omsat{\hat{\Omega}_{\rm sat}}

\def\nuisat{\nu_{\rm sat}}
\def\alphasat{\alpha_{\rm sat}}
\def\betai{\beta_0}

\def\bB{{\bf B}}
\def\vxi{{\vec\xi}}

\begin{document}
\title{R-Mode Oscillations and Spindown of Young Rotating
Magnetic Neutron Stars}
\author{Wynn C. G. Ho and Dong Lai}
\affil{Center for Radiophysics and Space Research, 
Department of Astronomy, Cornell University\\
Ithaca, NY 14853\\
E-mail: (wynnho, dong)@spacenet.tn.cornell.edu}

\begin{abstract}
Recent work has shown that a young, rapidly rotating neutron star 
loses angular momentum to gravitational waves generated by unstable 
r-mode oscillations. We study the spin evolution of a young, magnetic
neutron star including both the effects of gravitational radiation
and magnetic braking (modeled as magnetic dipole radiation).
Our phenomenological description of nonlinear
r-modes is similar to, but distinct from, that of Owen et al.~(1998)
in that our treatment is consistent with the principle of adiabatic 
invariance in the limit when direct driving and damping of the mode are
absent. We show that, while magnetic braking
tends to increase the r-mode amplitude by spinning down
the neutron star, it nevertheless reduces the efficiency 
of gravitational wave emission from the star. For $B\go 10^{14}
(\nus/300~{\rm Hz})^2$~G, where $\nus$ is the spin frequency,
the spindown rate and the gravitational waveforms are significantly 
modified by the effect of magnetic braking. 
We also estimate the growth rate of the r-mode due to electromagnetic
(fast magnetosonic) wave emission and due to \Alfven 
wave emission in the neutron star magnetosphere. The 
\Alfven wave driving of the r-mode becomes more important than the 
gravitational radiation driving when $B\go 
10^{13}(\nus/150~{\rm Hz})^3$~G; the electromagnetic wave driving 
of the r-mode is much weaker. 
Finally, we study the properties of local Rossby-\Alfven waves inside the
neutron star and show that the fractional change of the r-mode 
frequency due to the magnetic field is of order 
$0.5\,(B/10^{16}\,{\rm G})^2(\nus/100\,{\rm Hz})^{-2}$.
\end{abstract}


\section{Introduction}

Gravitational radiation drives unstable the r-modes of rotating neutron
stars (Andersson 1998; Friedman \& Morsink 1998) via the 
Chandrasekhar-Friedman-Schutz mechanism (``CFS instability''; Chandrasekhar
1970; Friedman \& Schutz 1978a,b).  Moreover, the r-mode
instability, due to current moment radiation, is so strong 
that viscous dissipations present in young neutron stars are 
insufficient to suppress the instability (Lindblom, Owen, \& Morsink 1998;
Andersson, Kokkotas, \& Schutz 1999; Lindblom \& Mendell 2000).
Owen et al.~(1998) modeled the spin evolution of a newly-formed, rapidly 
rotating neutron star and found that the growth of the r-mode 
causes the neutron star to spin down by losing angular momentum to
gravitational radiation and that the emitted gravitational waves 
can be a promising source for the LIGO/VIRGO gravitational wave detectors
currently under construction. 

It has been suggested that most neutron stars are born with relatively small
rotation rates (much less than the maximum break-up rate) and 
moderate magnetic fields of $10^{12}-10^{13}$~G (Spruit \& Phinney 1998).
These neutron stars are not efficient gravitational wave
emitters since the r-mode instability is suppressed by viscosity at slow
rotations. However, recent observations of soft gamma-ray repeaters and the
so-called ``anomalous x-ray pulsars'' (e.g., Vasisht \& Gotthelf 1997;
Kouveliotou et al.~1998, 1999) indicate that there exists a distinct 
population of neutron stars with much stronger magnetic fields, i.e.,
magnetars (Thompson \& Duncan 1993, 1995, 1996). The critical factor that
determines whether a proto-neutron star becomes an ordinary pulsar or a
magnetar may be its initial spin. It is conceivable that magnetars are
born with rapid rotations which facilitate dynamo actions that generate
their superstrong magnetic fields.
It is therefore possible that the r-mode instability may only be
relevant to proto-magnetars. 

In this paper we examine several issues related to the magnetic field
effects on the r-mode and on the spin evolution and gravitational wave 
emission of a newly-formed, rapidly rotating neutron star. An obvious 
issue is the magnetic braking of the neutron star due to magnetic dipole 
radiation. While magnetic braking is negligible compared to 
r-mode-driven spindown for canonical magnetic field strength ($\sim
10^{12}$~G) and rapid rotation, it becomes important for 
$B\go 10^{14}$~G and slower rotation. Magnetic braking can also drive/maintain
the r-mode instability indirectly by changing the background state of the
star according to the principle of adiabatic invariance. 
These effects of magnetic braking on the spin and r-mode evolution,
as well as the gravitational waveforms, are 
studied in \S2.  In \S3, we discuss several 
magnetic effects on r-modes, including r-mode instabilities driven by 
electromagnetic (fast magnetosonic) wave emission and by \Alfven wave
emission in the neutron star magnetosphere.  Section 3 also 
includes a discussion of how magnetic fields modify the r-mode
frequency and structure.  We summarize our results and discuss
their implications in \S4.

\section{Spin and R-mode Evolution}

\subsection{Basic Equations}

We first summarize the basic properties of r-modes needed for our
discussion [see Owen et al.~(1998) and references therein].  To leading
order in $\Omsb$ (the angular spin frequency of the neutron star),
the Eulerian velocity perturbation of the r-mode is of the form 
$\delta {\bf v}=\alpha\Omsb R(r/R)^l{\vec Y}_{lm}^B\,e^{i\omega t}$,
where $\alpha$ is the dimensionless amplitude of the mode, $R$ is the
neutron star radius, ${\vec Y}_{lm}^B=[l(l+1)]^{-1/2}r\nabla\times
(r\nabla Y_{lm})$ is the magnetic-type vector spherical harmonic of
order $l,m$. Throughout this paper, we consider only the dominant $l=m=2$ mode.
The angular frequency of the mode in the inertial frame and the
absolute value of the corresponding frequency in Hz are
$\omi = -4\Omsb/3$ and $\nui = |\omi|/(2\pi)=4\nu_s/3$,
where $\nu_s$ is the spin frequency 
(Papaloizou \& Pringle 1978; Provost, Berthomieu, \& Rocca 1981; Saio 1982).
The frequency of the gravitational wave emitted by the r-mode is also $\nu$.
We shall adopt the same neutron star model as in
Owen et al.~(1998), i.e., $M=1.4 M_\odot$, $R=12.53$~km,
and polytropic index $\Gamma=2$.  For such a neutron star, the damping
rate of the mode (i.e., in the linear regime, the mode amplitude evolves
as $e^{i\omi t - t/\tau}$) can be written as
${\tau}^{-1} = {\taugr}^{-1} + {\tauv}^{-1}$,
where $\taugr$ is the gravitational radiation timescale as given by
\be
\frac{1}{\taugr} =
-\frac{32\pi}{15^2}\lp\frac{4}{3}\rp^6 \frac{GMR^4}{c^7}\tJ\Omsb^6
= \frac{1}{\tilde\taugr} \Oms^6
=-\frac{1}{18.9\,{\rm s}} \lp\frac{\nus}{1~{\rm kHz}}\rp^6,
\label{eq:taugr}
\ee
and $\tauv$ is the viscous timescale as given by
\be
\frac{1}{\tauv} = \frac{T_9^{-2}}{\tilde{\tau}_{\rm S}}
 + \frac{T_9^{6}}{\tilde{\tau}_{\rm B}} \Oms^2.
\label{eq:tauv}\ee
Here ${\Oms}=\Omsb/\sqrt{\pi G\bar\rho}$,
$\tilde\taugr = -3.26$~s,
$\tilde{\tau}_{\rm S} = 2.52\times 10^8$~s,
$\tilde{\tau}_{\rm B} = 6.99\times 10^8$~s, $\tJ = 1.635\times10^{-2}$,
and $T=10^9 T_9$~K and $\bar{\rho}$ are the temperature and
mean density of the neutron star.

In the presence of r-mode oscillations, the neutron star will be 
differentially rotating. As gravitational radiation takes away angular 
momentum from the r-mode, the differential rotation is expected to
grow.\footnote{This is borne out by the exact solution of the evolution
of an unstable bar-mode in an incompressible Maclaurin spheroid as it 
evolves toward a
Dedekind ellipsoid driven by gravitational radiation reaction (Miller 1974;
Detweiler \& Lindblom 1977; Lai \& Shapiro 1995). See also Spruit (1999) and
Levin \& Ushomirsky (2000).}  Viscous forces tend to erase the differential 
rotation, i.e., to suppress the mode growth, by transferring angular momentum 
between the mode and the {\it mean} stellar rotation. In the absence of 
a precise nonlinear description of r-modes, we shall
derive the equations governing the evolution of the rotation
frequency and the r-mode amplitude using the phenomenological approach 
similar that of Owen et al.~(1998), although our equations
differ from theirs in an important aspect, even when $B=0$,
as we explain below. 

The total angular momentum of the neutron star can be decomposed into
the angular momentum due to a {\it mean} rigid rotation, 
$J = I\Omsb$, where $I$ is the moment of inertia, and the
canonical angular momentum of the mode,\footnote{The physical
angular momentum associated with the r-mode, defined as the change 
in angular momentum of the star due to the mode, generally differs
from $\Jc$ (Friedman \& Schutz 1978a; Levin \& Ushomirsky 2000).
This difference amounts to a proper definition of the mean rotation
rate $\Omsb$ for the differentially rotating star. The usefulness of $\Jc$
lies in the fact that in the absence of radiation and viscous dissipation,
$\Jc$ is conserved on account of the first-order perturbation equations,
and that $\Jc=-mI_{\rm c}$ is proportional to the wave action $I_{\rm c}$, 
an adiabatic invariant (Friedman \& Schutz 1978a); see discussion following
equation (\ref{eq:adot}).}
\be
\Jc = -\frac{3}{2}\tJ MR^2\Omsb\alpha^2. \label{eq:jc}
\ee
The canonical angular momentum of the mode can increase through
gravitational radiation and decrease by transferring angular
momentum to the star through viscosity.  Thus
\be
\frac{d\Jc}{dt} = -\frac{2}{\tau_{\rm GR}}\Jc - \frac{2}{\tau_{\rm V}}\Jc
+ N_{\rm other},
\label{eq:jcdot}
\ee
where $N_{\rm other}$ is the torque associated with other processes that can
drive the mode.  Section~3 discusses two possible mechanisms that
can contribute to $N_{\rm other}$, namely the driving of the mode
by electromagnetic radiation and by \Alfven wave
emission.  In this section, we consider driving only by
gravitational radiation, so that $N_{\rm other}=0$.

Another evolution equation is obtained from noting that the angular
momentum associated with the mean rigid rotation can increase
through a transfer from the mode due to viscosity and decrease
by magnetic braking.  Thus
\be
\frac{d\lp I\Omsb\rp}{dt} = \frac{2}{\tau_{\rm V}}\Jc - \frac{I\Omsb}{\taum},
\label{eq:jdot}
\ee
where $(-I\Omsb/\tau_M)$ is the magnetic braking torque and
$\taum$ is the corresponding timescale. Adopting the simplest 
magnetic dipole model, we have
\be
\frac{1}{\taum} = \frac{B^2R^6\Omsb^2}{6c^3I}
= \frac{1}{\tilde\taum} B_{14}^2 \Oms^2
= \frac{1}{1.21\times 10^5\,{\rm s}} B_{14}^2 \lp\frac{\nus}{1 {\rm kHz}}\rp^2.
\label{eq:taum}
\ee
Here $\tilde\taum = 6.74\times 10^4$~s, $B=10^{14}B_{14}$~G is the
neutron star magnetic field, and we have used $I=\tI MR^2$ with
$\tI=0.261$ for $\Gamma=2$ polytrope. Substituting equations~(\ref{eq:jc})
and (\ref{eq:taum}) into equation~(\ref{eq:jdot}) yields 
\be
\frac{d\Oms}{dt} = -2Q\frac{\Oms\alpha^2}{\tauv}
 - \frac{\Oms}{\taum}, \label{eq:spindot1}
\ee
where $Q = 3\tJ/2\tI = 9.40\times 10^{-2}$. From equations 
(\ref{eq:jcdot}) and (\ref{eq:spindot1})
we obtain the evolution equation for the mode amplitude,
\be
\frac{d\alpha}{dt} = -\frac{\alpha}{\taugr}
 -\frac{\alpha}{\tauv}\lp 1-\alpha^2 Q\rp
 + \frac{\alpha}{2\taum}.
 \label{eq:adot}
\ee

Note that our evolution equations (\ref{eq:spindot1}) and (\ref{eq:adot})
differ from those given in Owen et al.~(1998) even for $B=0$.
The difference arises because Owen et al.~(1998) derive one of their 
evolution equations by analyzing the rate of change of the mode energy
(in the rotating frame), namely
\be
\frac{dE_{\rm c}}{dt} = -2E_{\rm c}\lp\frac{1}{\taugr}+\frac{1}{\tauv}\rp,
\label{eq:edot}\ee
where $E_{\rm c} = \tJ MR^2\Omsb^2\alpha^2/2$.
This is in contrast to our equation~(\ref{eq:jcdot}) (with
$N_{\rm other}=0$) which is based on angular momentum considerations.
In the regime where the spin frequency does not change, the two
treatments give the same result for the evolution of the mode amplitude,
since $E_{\rm c}=(-\omr/2)J_{\rm c}$, where $\omr=2\Omsb/3$ is the 
mode frequency in the rotating frame. However, when $\Omsb$ changes,
the two treatments yield different results. We believe that
equation~(\ref{eq:edot}) is conceptually incorrect in the regime when both
$\Omsb$ and $\alpha$
are varying.  This can be most easily understood by considering 
a hypothetical situation where there is no gravitational radiation and
viscosity, but an external torque, such as magnetic braking,
slows the star down gradually. In such a situation, an initial 
wave, such as a r-mode, that propagates on the star will be amplified in
amplitude by $\alpha\propto \Omsb^{-1/2}$ according to the
conservation of wave action (an adiabatic invariant), which is simply 
proportional to $\Jc$. On the other hand, equation (\ref{eq:edot}) would
indicate $\alpha\propto \Omsb^{-1}$ for $\taugr^{-1}=\tauv^{-1}=0$,
which is clearly incorrect. Nevertheless, the difference between our 
evolution equations and those of Owen et al.~(1998) comes in at the level of 
$\alpha^2Q\simeq 0.1\alpha^2$, which is small unless the mode
becomes strongly nonlinear ($\alpha>1$). However, when the magnetic
braking effect is comparable to or greater than the gravitational
radiation effect, adopting equation (\ref{eq:edot}) would give a
significant error in the predicted behavior of the mode amplitude. 

There are three different stages in the evolution of the r-mode.
In the first stage, when the mode is still in the linear regime 
(with $\alpha\lo 1$), the mode amplitude grows according to equation
(\ref{eq:adot}), while the spin evolves according to equation
(\ref{eq:spindot1}). Note that for $\tau_M^{-1}\ll |\tau_{\rm GR}|^{-1}$,
the mode grows exponentially in this stage, but for 
$\tau_M^{-1}\go |\tau_{\rm GR}|^{-1}$, the mode growth can be significantly 
slower (see below). The second stage occurs when nonlinear 
effects prevent further growth of the mode. We adopt the same ansatz as
in Owen et al.~(1998), i.e., when the mode amplitude reaches 
$\alphasat =\sqrt{\kappa}\sim 1$, it will be held constant, so that
$d\alpha/dt=0$. In this stage, the spin evolution is then described by
\be
\frac{d\Oms}{dt} = \frac{2\Oms}{\taugr}\frac{\kappa Q}{1-\kappa Q}
 - \frac{\Oms}{\taum}\frac{1}{1-\kappa Q}.
\label{eq:spindot2}
\ee 
Finally, the third stage occurs when the star has been slowed down, 
and viscosity starts to damp out the mode. This stage begins 
when the right-hand side of equation (\ref{eq:adot}) becomes
negative. In the third stage, the evolution equations are again
given by equations~(\ref{eq:spindot1}) and (\ref{eq:adot}).

To evaluate the viscosities, we adopt the same temperature evolution 
of the neutron star, based on neutrino cooling via the modified URCA
process, as in Owen et al.~(1998), i.e.,
$T= 10^9 {\rm K}\,(t/1~{\rm yr})^{-1/6}$ for $t\go 10~{\rm s}$.
[For $t\lo 10$~s, $T\sim 10^{10}-10^{11}$~K, 
the star is opaque to neutrinos, and this bulk viscosity expression 
is invalid.]

\subsection{Results}

Integrating equations (\ref{eq:spindot1}), (\ref{eq:adot}), and
(\ref{eq:spindot2}) numerically, we can easily obtain the
mode amplitude and neutron star spin as functions of time.
Figure~\ref{fig:time} gives examples of such evolution for several different
magnetic field strengths: $B=0$~G, $10^{14}$~G, $10^{15}$~G,
$1.2\times 10^{15}$~G, and $2\times 10^{15}$~G.
We begin the star rotating at
$\Omsi=\hat{\Omega}_{\rm K}=2/3$, which corresponds
to $\nusi = 890$~Hz. We choose the initial mode amplitude $\alphai
=10^{-4}$ and the saturation amplitude $\alphasat = 1$.
In the absence of magnetic braking ($B=0$), the mode grows exponentially
during the first stage and reaches saturation at $\tsat \simeq 350$~s,
while the change in the spin frequency during this stage is on the order
of 0.1~Hz (Owen et al.~1998); the neutron star then spins down after 1 year to
$\nus \sim 90$~Hz. We see that for relatively weak magnetic fields
($B\lo 10^{14}$~G), magnetic braking affects the spin evolution 
only at later times, when the mode has saturated and when
$\taum^{-1}\propto B_{14}^2 \Omsb^2$ becomes comparable to
$|\taugr|^{-1}\propto \Omsb^6$. However, for sufficiently high magnetic
fields ($B\go 10^{15}$~G), the growth of the mode can be significantly delayed 
(e.g., $\tsat \simeq 1700$~s for $B=10^{15}$~G), and the star spins down
appreciably even in the first (linear) stage.

It is of interest to note that in the presence of magnetic braking, the mode
does not begin damping out even after the stellar spin has dropped to well
below $100$~Hz, the threshold spin frequency for the r-mode instability 
without magnetic braking. This comes about because magnetic braking
can drive (albeit slowly) and maintain the oscillation mode by changing
the background state, i.e., rotation, of the star in which the mode
lives, in accordance with the principle of adiabatic invariance. 
Indeed, equation (\ref{eq:adot}) indicates that the critical spin frequency
for r-mode growth is determined by the condition
\be
\frac{1}{\left|\taugr\right|}-\frac{1}{\tauv}+\frac{1}{2}\frac{1}{\taum} > 0,
\label{eq:crit}
\ee
where we have set $\lp 1-\alpha^2 Q\rp\simeq 1$. 
Figure~\ref{fig:temp} depicts the critical spin frequencies as functions
of the neutron star temperature for several different values of $B$. 
The evolutionary tracks are also shown for comparison. 
It is evident that magnetic fields decrease the requisite critical 
spin frequencies for mode growth by countering the viscous damping.
At late times, when the neutron star has cooled, it is the shear
viscosity that damps the mode.  Without magnetic braking, this
viscosity prevents the growth of the mode when
$\nus < 65\,T_9^{-1/3}$~Hz. However, magnetic braking can sustain 
the growth of the mode if $\nus \go 31\,T_9^{-1} B_{14}^{-1}$~Hz. 

The effects of magnetic braking depend on both 
$B$ and the initial spin $\Omsib$ of the neutron star.
The critical relevant parameter is the ratio of the 
magnetic braking timescale and the gravitational radiation timescale,
\be
\beta\equiv \frac{\taum}{|\taugr|}
= \frac{64\pi}{75} \lp\frac{4}{3}\rp^6
\frac{GM^2}{c^4}\tI\tJ\frac{\Omsb^4}{B^2}
= 0.64\,B_{14}^{-2} \lp\frac{\nus}{100~{\rm Hz}}\rp^4.
\ee
If the initial spin $\Omsib$ is such that $|\taugr|^{-1}\gg
\tauv^{-1}$ or $|\taum|^{-1}\gg \tauv^{-1}$, we can neglect the
viscous term in equations~(\ref{eq:spindot1}) and (\ref{eq:adot}).
We then find that $\tsat$ and $\Omsat$, the time and rotation rate
at the saturation point, respectively, are determined by 
\be
1=-{1\over 2\gamma}\ln\xsat +{\betai \over 4\gamma}(1-\xsat^4),\qquad
\ysat=\tsat/\tsato={\betai\over 2\gamma}(1/\xsat^2-1),
\ee
where $\betai$ is the initial $\beta$,
$\xsat = \Omsat/\Omsi$, $\gamma=\ln\lp\alphasat/\alphai\rp$,
and $\tsato = \left|\taugr\right|_0 \ln\lp\alphasat/\alphai\rp$
is the time to saturation without magnetic braking.
For $\betai/\gamma\gg 1$, we find
\be
\xsat  =  1-{\gamma\over\betai}-{\gamma^2\over\betai^2}\left({3\over 2}-{1
\over 2\gamma}\right),\qquad
\ysat  =  1+\left(3-{1\over 2\gamma}\right){\gamma\over\betai}.
\ee
Figure~\ref{fig:beta} shows the general results for $\xsat$ and
$\ysat$ with $\gamma=\ln 10^4$ and $\ln 10^8$.
It is clear that magnetic braking increases both $\tsat$ and
$\Delta\Omsat= \Omsi-\Omsat = \lp 1-\xsat \rp \Omsi$.
Note that in the $B=0$ limit, $\Delta\Omsat$ is determined
by viscosity and is of order $10^{-4}$; the expressions above are valid
as long as the resulting $\Delta\Omsat$ is greater than the zero-field
limit. From Fig.~\ref{fig:beta}, we see that the linear
evolution will be significantly affected when 
\be
\betai \lo 5\gamma\simeq 50 \quad \Longleftrightarrow \quad
\nusi \lo 300\,\left({\gamma\over 10}\right)^{1/4} 
B_{14}^{1/2}~{\rm Hz} \label{eq:beta0}.
\ee
The transition is quite dramatic. For example, we see from Fig.~\ref{fig:time}
that the time it takes for the mode to saturate is $\tsat\sim 1700$~s
for $B=10^{15}$ when $\nu_{s0}=890$~Hz.  But for $B=1.2\times 10^{15}$~G,
saturation has not been reached even after $t=1$~yr. This will 
significantly affect the gravitational wave signals (see \S2.3).
In the second (saturation) stage, the spindown rate is given by 
[see eq.~(\ref{eq:spindot2})]
\be
{1\over\nus}\frac{d\nus}{dt}\simeq -\frac{\kappa}{3~{\rm yr}}
\lp\frac{\nus}{100~{\rm Hz}}\rp^6 \lp 1 + \frac{1}{2\kappa Q \beta}\rp.
\ee
Thus, even for $\betai\gg 50$, so that the linear evolution is not affected,
magnetic braking can still affect the evolution in the 
the second (saturation) stage, when 
\be
\beta \lo {1\over 2\kappa Q}\simeq 5
 \quad \Longleftrightarrow \quad
\nus \lo 170\,\kappa^{-1/4} B_{14}^{1/2}~{\rm Hz}.
\label{eq:beta5}\ee

\subsection{Gravitational Waveforms}

The rate of angular momentum loss due to the r-mode can be written as
\be
\lp\frac{d\Jc}{dt}\rp_{\rm GR} = - \frac{2}{\taugr}\Jc
= 4\pi d^2 \frac{m}{\omi} \frac{\omi^2}{16\pi}
\left<h_+^2 + h_\times^2\right>,
\label{eq:jdotgr2}
\ee
where $h_+$ and $h_\times$ are the strain amplitudes for the two
polarizations of the gravitational wave and $d$ is the distance
to the source (Shapiro \& Teukolsky 1983).
Following Owen et al.~(1998), we define the averaged
gravitational wave amplitude $h(t)$ by
$\lb h(t)\rb^2 = (3/10)\left<h_+^2 + h_\times^2\right>$.
Using equations~(\ref{eq:jc}) for $\Jc$ and (\ref{eq:taugr}) for
$\taugr$, we find
\be
h(t)  =  \frac{256}{45}\sqrt{\frac{\pi}{30}}\frac{GMR^3}{c^5}\tJ
 \frac{\alpha\Omsb^3}{d} 
=  1.8\times10^{-24} \alpha
\lp\frac{\nus}{1~{\rm kHz}}\rp^3\lp\frac{20 {\rm Mpc}}{d}\rp. \label{eq:h2}
\ee
The bottom panel of Fig.~\ref{fig:time} plots $h(t)$ as a function of
time for several different magnetic field strengths, with 
the distance to the source $d = 20$~Mpc.
As expected, the effect of magnetic braking is to decrease the
gravitational wave amplitude and shorten the duration of 
gravitational wave emission.

To examine the detectability of gravitational waves, we 
define the characteristic gravitational wave amplitude $h_{\rm c}$ via
\be
h_{\rm c}(\nui) \equiv \nui\th(\nui)=\nui\, {h(t)\over |d\nui/dt|^{1/2}},
\label{eq:hhft}
\ee
where $\th$ is the Fourier transform of the gravitational wave amplitude.
In equation~(\ref{eq:hhft}), $d\nui/dt$ can be obtained from
equations~(\ref{eq:spindot1}) and (\ref{eq:spindot2}) for the three stages
of spindown evolution. Figure~\ref{fig:hc} shows 
$h_{\rm c}(\nu)$ for two different initial spin frequencies ($\nu_{s0}=890$~Hz
and $400$~Hz), each with several different values of $B$.  The other parameters
are the same as in Fig.~\ref{fig:time} and Fig.~\ref{fig:temp}.
For relatively weak magnetic fields, the vertical spikes seen in 
Fig.~\ref{fig:hc} result from the fact that when $\betai\gg 50$, the spin
frequency, and hence $\nui$, is essentially constant before
saturation.\footnote{Recent work by Levin \& Ushomirsky (2000) suggests
that the spindown evolution may be unrelated to mode saturation.  The mode
frequency in this case may not be constant before saturation, and thus
the vertical spikes in Fig.~\ref{fig:hc} would be broadened.  The
details, however, are uncertain at this time.}
For higher magnetic field (smaller $\betai$), the spikes are broadened,
and the width of the spike, $\Delta\nu=(2/3\pi)(1-\xsat)\Omsib$, 
becomes significant for $\betai\lo 50$ [see eq.~(\ref{eq:beta0}) and
Fig.~\ref{fig:beta}]. Near saturation, the
gravitational wave amplitude is
\be
h_{\rm c}(\nui=\nuisat)
= 4.9\times 10^{-21}\lp\frac{\nuii}{1~{\rm kHz}}\rp^{-1/2}
\gamma^{1/2} \xsat^{5/2} \ysat^{1/2}
\lp 1+118\frac{\gamma \ysat}{\kappa\beta{\Omsi}^2}\rp^{-1/2}.
\label{eq:hc2}
\ee
For small $B$, we have $\xsat=\ysat=1$, and equation~(\ref{eq:hc2}) 
reduces to the zero-field result.
In the saturated regime, we have
\be
h_{\rm c}(\nu) \simeq  5.8\times 10^{-22}\lp\frac{\nui}{1~{\rm kHz}}\rp^{1/2}
\lp 1+\frac{1}{2\kappa Q\beta}\rp^{-1/2}.
\label{eq:hcsat}\ee
Thus for $\beta\lo 1/(2\kappa Q)\simeq 5$ [see eq.~(\ref{eq:beta5})],
the characteristic amplitude is reduced from the zero-field value. 

In Fig.~\ref{fig:hc}, we also compare $h_{\rm c}(\nu)$ 
with the expected RMS noise spectra $h_{\rm rms}
=\sqrt{\nui S_h(\nui)}$ of three (initial,
enhanced, and advanced) generations of LIGO, where
$S_h\lp\nui\rp$ is the power spectral density of the detector
strain noise.  The expressions for $S_h(\nu)$ are given in 
Owen et al.~(1998). We see that the signal-to-noise ratio,
proportional to $\nu^{-1}(h_{\rm c}/h_{\rm rms})^2$ integrated over the
frequency band, can be significantly reduced by the magnetic braking effect. 

\setcounter{equation}{0}
\section{Effects of Magnetic Fields on R-modes}

Magnetic braking discussed in \S2 only indirectly affects the r-mode 
instability. In this section, we discuss several direct effects 
of magnetic fields on r-mode oscillations and their instabilities.

\subsection{Driving of R-Mode by Electromagnetic Radiation}

R-mode oscillations in the neutron star perturb the magnetic field anchored
on the star, and this leads to electromagnetic radiation. Just like
gravitational radiation, the electromagnetic radiation also drives r-modes
unstable via the CFS mechanism. Consider the $l=m=2$ 
r-mode oscillation with surface Lagrangian displacement
\be
{\vec \xi} = \xi_0\left(0,\frac{1}{\sin\theta}
\frac{\partial}{\partial \phi},-\frac{\partial}
{\partial \theta}\right)Y_{22}e^{-i\omi t}, \label{eq:xir}
\ee
where $\omega=4\Omsb/3$ (we shall adopt the convention $\omega>0$ in this
section and in Appendix A). The power of electromagnetic radiation 
associated with this oscillation is given by (see Appendix A)
\be
P_{\rm EM}\simeq {cR^2\over 108\pi}\left({\omega\xi_0\over c}B\right)^2
\left({\omega R\over c}\right)^4. \label{eq:pem}
\ee
Since the mode energy (in the inertial frame) is 
$E_{\rm c}^{\rm (in)}=-(9/56\pi)M\omi^2\xi_0^2$ for a uniform star,
the timescale of EM driving is then
\be
\frac{1}{\tauem} = -{7\over 3^5}\lp\frac{B^2 R^3}{Mc^2}\rp
\left({\omega R\over c}\right)^3\omega
= -\frac{1}{(3.8\times 10^6~{\rm s})}M_{1.4}B_{14}^2\Oms^4.
\ee
Comparing this with $\tau_{\rm GR}^{-1}=-(4^8/3^7)(GMR^4\Omsb^6/350c^7)$
(for a uniform neutron star), we find
\be
\frac{1/\tauem}{1/\taugr}
\simeq 0.47\,M_{1.4}^{-2} R_{10}^2 B_{16}^2
\lp\frac{100 {\rm Hz}}{\nus}\rp^2, \label{eq:tauem}
\ee
where $M_{1.4}$ is the neutron star mass in units of $1.4 M_\odot$,
$R_{10}$ is the radius in units of 10~km, and
$B_{16}$ is the magnetic field in units of $10^{16}$~G.
Clearly, electromagnetic driving of r-mode is negligible unless the neutron
star has $B\go 10^{16}$~G.

Since a real neutron star possesses a magnetosphere, whose plasma frequency 
is much greater than the typical rotation frequency, 
it is unlikely that electromagnetic radiation can be emitted from the star. 
However, as pointed by Blaes et al.~(1989), in a relativistic
plasma, fast magnetosonic waves have the same dispersion relation 
as electromagnetic waves, and their emission is therefore 
analogous to that of electromagnetic waves.

\subsection{Driving of R-Mode by \Alfven Wave Emission}

The shaking of magnetic field lines by r-mode oscillations also results
in \Alfven wave emission to the magnetosphere. The torsional \Alfven waves 
carry away energy and angular momentum from the r-mode and thus
drive it unstable. Because of the uncertainties in the 
magnetosphere physics, here we shall be content with an order-of-magnitude
estimate of the \Alfven wave power. 
In a relativistic plasma, the group velocity of \Alfven waves is
close to the speed of light and is along the magnetic field line. 
A fluid displacement $\xi_0$ on the neutron star surface produces
a kink in the attached field line which in turn attempts to straighten 
up by launching \Alfven waves.  In one oscillation cycle, the signal
travels a distance $\sim c/\omega$. Therefore the amplitude of the \Alfven
wave is of order $\delta B\sim (\omega\xi_0/c)B$.  Note that this is
smaller than the magnetic field perturbation, of order $(\xi_0/R)B$, 
inside the stellar surface. The power radiated by the \Alfven
waves is then $P_{\rm A} \sim [(\delta B)^2/(8\pi)](4\pi R^2)c
=(R\omega\xi_0 B)^2/(2c)$. An improved calculation,
taking into account various geometric factors, yields
\be
P_{\rm A}\sim {c\over 8\pi}\int_0^\pi\!\left|{\vec\xi}(R)\times{\bf B}\right|^2
{\omi^2\over c^2}\, 2\pi R^2\,|\hat B\cdot\hat r|\,
\sin\,\theta\,d\theta
\simeq 0.04\,cR^2\left({\omega\xi_0\over c}B\right)^2,
\label{eq:alfven}\ee
where we have assumed the field to be dipolar outside the stellar
surface.\footnote{Note that the power radiated by \Alfven waves scales
as $\omi^2$ while the power emitted by electromagnetic (fast magnetosonic)
waves scales as $\omi^6$ [see eq.~(\ref{eq:pem}].  The weak coupling
of the magnetosonic waves can be understood in the following way
[see Blaes et al. (1989) for a similar discussion on elastic waves].
Prior to reaching the stellar surface, the oscillation mode has
the dispersion relation $\omi\sim kv$, where $v\sim\omi R$.
The magnetosonic and \Alfven
waves in the magnetosphere have the dispersion relation $\omi=kc$
and $\omi=kc\cos\psi$, respectively, where $\psi$ is the angle
between ${\bf k}$ and ${\bf B}$.  From the continuity of the
projection of the wave vectors onto the incident boundary,
$k_i\sin\theta_i=k_t\sin\theta_t$ (Snell's law), where $\theta_i$
and $\theta_t$ are the angles of incidence and transmission,
we have $\sin\theta_t=(c/v)\sin\theta_i$ and
$\sin\theta_t=(c/v)\cos\psi\sin\theta_i$
for the magnetosonic and \Alfven waves, respectively.  Since $c\gg v$,
the oscillation modes must have a very small angle of incidence
in order for them to be transmitted (coupled) to the magnetosonic
waves, otherwise the mode is reflected at the boundary.  On the
other hand, the additional degree of freedom allowed by $\cos\psi$
permits a wide range for the angle of incidence of the oscillation
modes in the case of \Alfven waves, so that the modes are more easily
coupled to \Alfven waves, with the only restriction that ${\bf k}$ be
almost orthogonal to ${\bf B}$.}
Obviously equation~(\ref{eq:alfven}) should be considered as
an estimate since we do not solve for the global field perturbation
self-consistently. The timescale of \Alfven wave driving is then
\be
\frac{1}{\taua} \sim -0.4 
\lp\frac{B^2 R^3}{Mc^2}\rp\lp\frac{c}{R}\rp
= -\frac{1}{2\times 10^4~{\rm s}} M_{1.4}^{-1} R_{10}^2 B_{14}^2.
\ee
Comparing the \Alfven wave power to the gravitational 
radiation power, we have
\ba
\frac{1/\taua}{1/\taugr}
\sim  10^3\,M_{1.4}^{-2} R_{10}^{-2} B_{14}^2
\lp\frac{100 {\rm Hz}}{\nus}\rp^6.
\label{eq:taua}\ea
Thus, when the spin frequency
$\nus\lo 320$~Hz~$B_{14}^{1/3}=150$~Hz~$B_{13}^{1/3}$,
\Alfven wave emission becomes more important than gravitational
radiation in driving the r-mode instability. 

We note that our calculation for the \Alfven wave power, equation 
(\ref{eq:alfven}), includes the energy transmitted through both
the open and closed magnetic field lines.  The energy transmitted
through the open field lines will be dissipated at infinity; 
the energy associated with the closed field lines is
less clear.  Also, only the coherent part of the \Alfven wave carries
away canonical mode energy, thereby driving the mode.
In the absence of a complete understanding of the global properties of
the magnetosphere, we cannot definitively calculate the mode driving/damping
rate due to \Alfven wave emission. If we include only the 
power radiatied through the open field lines, then
equations~(\ref{eq:alfven})-(\ref{eq:taua}) must be modified.
The integral in equation~(\ref{eq:alfven}) is instead evaluated from
$\theta = 0$ to $\theta = \theta_{\rm cap} = \sin^{-1}\sqrt{\Omsb R/c}$.
This results in the power and driving rate of \Alfven wave radiation
being multiplied by a factor of
$\sim 4\sin^4\theta_{\rm cap} \simeq 4(\Omsb R/c)^2 
\simeq 2\times 10^{-3}R_{10}^2(\nus/100~{\rm Hz})^2$.

An important caveat in our calculations of $\tau_{\rm A}$ and
$\tau_{\rm EM}$ is the neglect of neutron star crust, which
is expected to form for $T\lo 10^{10}$~K. The r-mode oscillation 
in the fluid core may not penetrate the crust. As a result, the 
r-mode-induced perturbations to the magnetosphere are greatly reduced
except for hot neutron stars.  

\subsection{Rossby-\Alfven Waves}

A r-mode oscillation perturbs the magnetic field inside the star,
and the resulting magnetic stress acts as a restoring force for the
oscillation. If the magnetic field is sufficiently strong so
that the \Alfven speed becomes comparable to the phase
velocity of the r-mode, the oscillation frequency and eigenfunction will be
significantly modified by the magnetic field. In essence, the 
oscillation does not correspond to a pure Rossby wave but contains
a mixture of Rossby wave and \Alfven wave characteristics.

While a global analysis of the Rossby-\Alfven mode of a rotating magnetic
star is beyond the scope of this paper, it is instructive
to examine the properties of local Rossby-\Alfven waves. 
We consider fluid perturbation ${\vec\xi}$ in the neighborhood of 
$(r,\theta,\phi)$ inside a fluid star which rotates at angular frequency
$\Omsb$ around the $z$-axis. It is convenient to 
set up a local Cartesian coordinate $(x_1,x_2,x_3)$ so 
that $dx_1=rd\theta,\,dx_2=r\,\sin\theta d\phi,\,dx_3=dr$.
We consider the simplest case in which the unperturbed magnetic field
$B$ is along the radial direction, and ${\vec \xi}$ has only $x_1,x_2$ 
components. The oscillation involves no density perturbation and
thus satisfies $\nabla\cdot\vxi=0$.
The perturbed magnetic field is given by
\be
\delta\bB=\nabla\times (\vxi\times\bB)=B{\partial\over\partial x_3}\vxi,
\ee
where we have neglected the curvature of the unperturbed magnetic field. 
The $x_1,x_2$ components of Euler's equation in the rotating frame can 
be written as
\ba
&&{\partial^2\xi_1\over\partial t^2}=-{1\over\rho}{\partial\delta P
\over\partial x_1}+2\Omega_r{\partial\xi_2\over\partial t}
+{B^2\over 4\pi\rho}{\partial^2\xi_1\over\partial x_3^2} \label{eq:xi1}\\
&&{\partial^2\xi_2\over\partial t^2}=-{1\over\rho}{\partial\delta P
\over\partial x_2}-2\Omega_r{\partial\xi_1\over\partial t}
+{B^2\over 4\pi\rho}{\partial^2\xi_2\over\partial x_3^2},\label{eq:xi2}
\ea
where $\Omega_r=\Omsb\,\cos\theta$, and the last terms of the equations
represent the magnetic force per unit mass given by ${\bf F}_B=(\bB\cdot
\nabla)\delta\bB/(4\pi\rho)$.  (The displacement current in the Maxwell 
equation is neglected since the \Alfven speed inside the star is much less
than the speed of light.) We can eliminate the pressure perturbation 
$\delta P$ from equations~(\ref{eq:xi1}) and (\ref{eq:xi2}), and then set
$\vxi\propto e^{i\omr t+ik_1x_1+ik_2x_2+ik_rx_3}$, where
$\omr$ is the mode frequency in the rotating frame.
Combining the resulting equation with 
$\nabla\cdot\vxi=i(k_1\xi_1+k_2\xi_2)=0$,
we obtain the dispersion relation
\be
\omr\left(1-{\va^2k_r^2\over\omr^2}\right)=-2\left({\partial\Omega_r
\over\partial x_1}\right){k_2\over k_1^2+k_2^2}\equiv\omr_0,
\label{eq:disp}\ee
where $\va=(B^2/4\pi\rho)^{1/2}$ is the \Alfven speed. 
For $B=0$, this is simply the dispersion relation for local 
Rossby waves. Indeed, with $\partial\Omega_r/\partial x_1
=-\Omsb\,\sin\theta/r$, $k_2=m/(r\,\sin\theta)$ and 
$k_1^2+k_2^2=l(l+1)/r^2$, the right-hand-side of
equation~(\ref{eq:disp}) becomes $\omr_0=2m\Omega/[l(l+1)]$,
the r-mode frequency of order $l,m$. For $B\neq 0$, the
Rossby-\Alfven wave frequency is given by
\be
\omr={\omr_0\over 2}\pm {1\over 2}
\left(\omr_0^2+4\oma^2\right)^{1/2},
\ee
where $\oma=\va k_r$ is the local \Alfven angular frequency
\be
\oma\simeq 300\,(k_rR)B_{16}\rho_{14}^{-1/2}R_{10}^{-1}~{\rm s}^{-1}.
\ee
For $\omr_0\gg\oma$, the change of the r-mode frequency,
with the $l=m=2$ mode and $k_rR\sim 1$, is of order
\be
{\Delta\omr\over\omr_0}={\oma^2\over\omr_0^2}
\simeq 0.5\,B_{16}^2\rho_{14}^{-1}R_{10}^{-2}\left({100\,{\rm Hz}\over
\nus}\right)^2.
\ee
For $\Delta\omr/\omr_0\ll 1$, the oscillation manifests
itself as Rossby waves in the bulk interior of the star, and the 
magnetic energy associated with the oscillation is negligible. 
But near the stellar surface, where $\oma\gg \omr_0$, the 
oscillation assumes the \Alfven wave character. 

\section{Discussion}

We have studied the combined effects of magnetic braking and 
gravitational radiation generated by unstable r-mode oscillations
on the spin evolution of a young, rapidly rotating neutron star.
Since the gravitational radiation timescale of r-mode scales as
$\nus^{-6}$, where $\nus$ is the spin frequency, while
the magnetic braking time scales as $(B\nus)^{-2}$, it is obvious
that the magnetic braking becomes important only for sufficiently large
$B$ and small $\nus$ [see eqs.~(2.15) and (2.17)]. We have
examined the characteristics of the gravitational waveforms for different
magnetic field strengths. As discussed in \S1, it is possible that 
young rapidly rotating neutron stars are necessarily endowed 
with superstrong magnetic fields ($B\go 10^{14}$~G); thus one (an optimist)
might suggest that the spectral features of the gravitational waves 
can be used to probe the initial magnetic fields and spins
of these neutron stars. 

Magnetic fields can affect the r-mode oscillation in different ways. 
We have discussed a few of these effects in this paper (see \S3).
For sufficiently large $B$ and slow rotation, the driving mechanisms
of the r-mode due to electromagnetic (fast magnetosonic) wave
emission and \Alfven wave emission can compete with the driving due to
gravitational radiation.
In an interesting paper, Rezzolla et al.~(2000) suggested that r-mode 
oscillations can wind up the stellar magnetic field due to differential 
drifts of fluid elements,\footnote{We note that in the case of an
incompressible Maclaurin spheroid, where the nonlinear $m=2$ f-mode
is exactly known and is described by a Riemann S-ellipsoid, there is 
no differential drift and therefore no secular winding of field lines.}
and this serves as a damping mechanism for the r-mode.
The damping rate estimated by Rezzolla et al.~(2000), if correct, is about
$(c/\Omsb R)$ times larger than our estimated growth rate due to \Alfven
wave emission (assuming that the toroidal magnetic field is of the same
order as the poloidal field). It is of interest 
to note that their estimate of the damping rate,
$\gamma$, is of the same order of magnitude 
as the change in the mode frequency due to the
magnetic field, i.e., $\gamma\sim \Delta\omr\sim
\oma^2/\omr_0$ (see \S 3.3).

Bildsten \& Ushomirsky~(2000) have shown that the presence of a solid
crust in the neutron star can strongly enhance the viscous damping of the
r-mode.\footnote{The boundary layer analysis of Bildsten \& Ushomirsky~(2000)
hinges upon the existence of a sharp (width less than a few cm) transition
between the liquid core and the inner crust, i.e., a first order phase
transition. While this may well be true (Baym et al.~1971), one cannot 
be completely sure. Note that the pressure scale height of this region is
of order $10^4$~cm. It is also not clear how rigid the the core-crust
transition region is, given that the strong interaction plays a more 
dominant role than the electrostatic interaction in determining the
properties of the medium, contrary to the situation in 
a classical one-component plasma for which the original melting 
criterion is obtained.  Nevertheless, it clear that the crust
can enhance the viscous dissipation, although one might consider 
the estimate of Bildsten \& Ushomirsky~(2000) as an upper limit.
Recent calculations by Andersson et al. (2000) suggest that
Bildsten \& Ushomirsky (2000) overestimated the strength of the damping.
Owen (1999) and Andersson et al. (2000) also point out that
the energy released by a mode that is already present may delay
the formation of the crust, thus negate the large viscous dissipation
due to the boundary layer, and allow the spindown evolution to proceed
by the method described in our paper.}
For viscous damping $10^5$ times stronger [as suggested by
Bildsten \& Ushomirsky (2000)], the critical spin frequency below which 
the mode begins to damp is given by $\nus \simeq 440\,T_9^{-1/3}$~Hz
for low magnetic fields. 
It is of interest to note that when the magnetic field is sufficiently strong,
the mode does not decay until $\nus \lo 100\,T_9^{-1} B_{16}^{-1}$~Hz due to
the indirect driving of the mode by magnetic braking 
[see eq.~({\ref{eq:crit})]. Nevertheless, for $B \go 10^{15}$~G, 
the strength of the gravitational waves is suppressed because magnetic braking
dominates over gravitational radiation in the spindown, as shown in
Fig.~\ref{fig:hc}.

\acknowledgments

Some of the calculations presented in this paper were done in the summer
of 1998 while D.L. was at the Aspen Center for Physics. He thanks
the participants of JENAM'99 ``Gravitational Wave Astronomy'' Workshop
(September 1999, Toulouse, France) for their interest in this work, which
prompted us to write this paper. He also acknowledges helpful discussion 
with Ira Wasserman, who has independently considered some of the 
issues addressed in this paper.  Finally, we thank Yuri Levin for
useful comments.  This work is supported by 
NASA Grant NAG 5-8484 and an Alfred P. Sloan Foundation fellowship to D.L.

\appendix

\setcounter{equation}{0}
\section{R-Mode Driving by Electromagnetic Radiation}

Consider a star with a uniform magnetic field.
We set up a coordinate system $(x,y,z)$ centered on the star with
the $z$-axis along ${\bf \Omsb}$, such that ${\bf \Omsb} = \Omsb{\hat z}$.
Let the angle between the $z$-axis and the magnetic field axis be $\alpha$.
The magnetic field is then given in spherical coordinates as
\ba
{\bf B} & = & B_0 \lb \cos\alpha \, \cos\theta
 + \sin\alpha \, \sin\theta \, \cos \lp \phi - \Omsb t \rp \rb \hat{r} \nonumber \\
 && + B_0 \lb - \cos\alpha \, \sin\theta
 + \sin\alpha \, \cos\theta \, \cos \lp \phi - \Omsb t \rp \rb \hat{\theta} \nonumber \\
 && - B_0 \lb \sin\alpha \, \sin \lp \phi - \Omsb t \rp \rb \hat{\phi}.
\ea
The Lagrangian displacement of the r-mode at the stellar surface 
is given by equation (\ref{eq:xir}), so that the velocity perturbation is
\be
\delta{\bf v} = -i\omi \xi_0\lp\frac{2i}{\sin\theta} Y_{22}\hat{\theta}
-\frac{\partial Y_{22}}{\partial\theta}\hat{\phi}\rp e^{-i\omi t}.\label{eq:vr}
\ee
Note that, for the discussion here, we have changed the sign of
$\omi$ so that $\omi = 4\Omsb/3 > 0$.  The star
then has an induced electric field
${\bf E}=-\lp\delta{\bf v}/c\rp\times{\bf B}$,
which possesses three components with different frequencies.  These are
\ba
{\bf E}_{\rm mode,1}&=&-\frac{i\omi \xi_0 B_0}{c}\,\cos\alpha
\,\cos\theta \lp\frac{\partial Y_{22}}{\partial\theta} {\hat\theta}
+ \frac{2i}{\sin\theta}Y_{22} {\hat\phi} \rp e^{-i\omi t}
\label{eq:e1} \\
{\bf E}_{\rm mode,2}&=&-\frac{i\omi \xi_0 B_0}{2c}\,\sin\alpha
\lp\sin\theta\frac{\partial Y_{22}}{\partial\theta} {\hat\theta}
+ 2iY_{22} {\hat\phi} \rp e^{-i\phi-i\lp\omi-\Oms\rp t}
\label{eq:e2} \\
{\bf E}_{\rm mode,3}&=&-\frac{i\omi \xi_0 B_0}{2c}\,\sin\alpha
\lp\sin\theta\frac{\partial Y_{22}}{\partial\theta} {\hat\theta}
+ 2iY_{22} {\hat\phi} \rp e^{-i\phi-i\lp\omi+\Oms\rp t}.
\label{eq:e3}
\ea
Thus the $l=m=2$ mode gives rise to electromagnetic waves with
frequencies $\omi,\omi+\Omsb$, and $\omi-\Omsb$. 
Using the method and notation described of Jackson (1975), we can obtain
coefficients $a_{\rm E}(l,m)$ and $a_{\rm M}(l,m)$
which specify the electric $(l,m)$ multipoles and magnetic $(l,m)$
multipoles. For the electromagnetic radiation corresponding to (\ref{eq:e1}),
the nonzero coefficients, up to $l=3$, are
\ba
a_{\rm M}(22) & = & -\frac{2}{\sqrt{6}} \frac{\lp kR\rp^3}{3i}
\frac{i\omi \xi_0 B_0}{c}\cos\alpha \\
a_{\rm E}(32) & = & \frac{1}{\sqrt{12}}\lp\sqrt{7}+\frac{1}{\sqrt{7}}\rp
\frac{\lp kR\rp^5}{45i} \frac{i\omi \xi_0 B_0}{c}\cos\alpha,
\ea
where $k=\omi/c$.
For the electromagnetic radiation corresponding to (\ref{eq:e2}),
\ba
a_{\rm E}(11) & = & - \frac{1}{\sqrt{2}}\lp\sqrt{5}+\frac{1}{\sqrt{5}}\rp
\frac{\lp kR\rp^3}{i} \frac{i\omi \xi_0 B_0}{2c}\sin\alpha \\
a_{\rm M}(21) & = & -\frac{2}{\sqrt{6}} \frac{\lp kR\rp^3}{3i}
\frac{i\omi \xi_0 B_0}{2c}\sin\alpha \\
a_{\rm E}(31) & = & \frac{3}{\sqrt{12}}\sqrt{\frac{2}{35}}
\frac{\lp kR\rp^5}{45i} \frac{i\omi \xi_0 B_0}{2c}\sin\alpha,
\ea
where $k=\lp\omi-\Omsb\rp/c$.
For the electromagnetic radiation corresponding to (\ref{eq:e3}),
\ba
a_{\rm E}(33) & = & -\frac{8}{7}\sqrt{\frac{7}{2}}
\frac{\lp kR\rp^5}{45i} \frac{i\omi \xi_0 B_0}{2c}\sin\alpha,
\ea
where $k=\lp\omi+\Omsb\rp/c$.
To leading order in $\lp\omi R/c\rp$, the contribution to the power
from (\ref{eq:e1}) and (\ref{eq:e2}) are, respectively,
\ba
P_1 & = &\frac{1}{108\pi}\frac{R^6}{c^5}\omi^6\xi_0^2 B_0^2\lp\cos\alpha\rp^2 \\
P_2 & = &\frac{61}{34560\pi}\frac{R^6}{c^5}\omi^6\xi_0^2B_0^2\lp\sin\alpha\rp^2.
\ea
The power from (\ref{eq:e3}) is negligible.
Comparing the total power emitted from electromagnetic radiation to the
rate of energy loss due to gravitational radiation gives the timescale
for electromagnetic driving of the mode in terms of the timescale for
gravitational driving, namely
\be
\frac{1/\tauem}{1/\taugr} = 0.47\,M_{1.4}^{-2} R_{10}^2
B_{16}^2 \lp\cos^2\alpha + \frac{61}{80}\sin^2\alpha\rp
\lp\frac{100 {\rm Hz}}{\nus}\rp^2. \label{eq:tauem0}
\ee


\label{lastpage}

\begin{figure*}
\epsfxsize=17cm
\epsfysize=17cm
\hbox{\hskip -0 truecm \epsffile{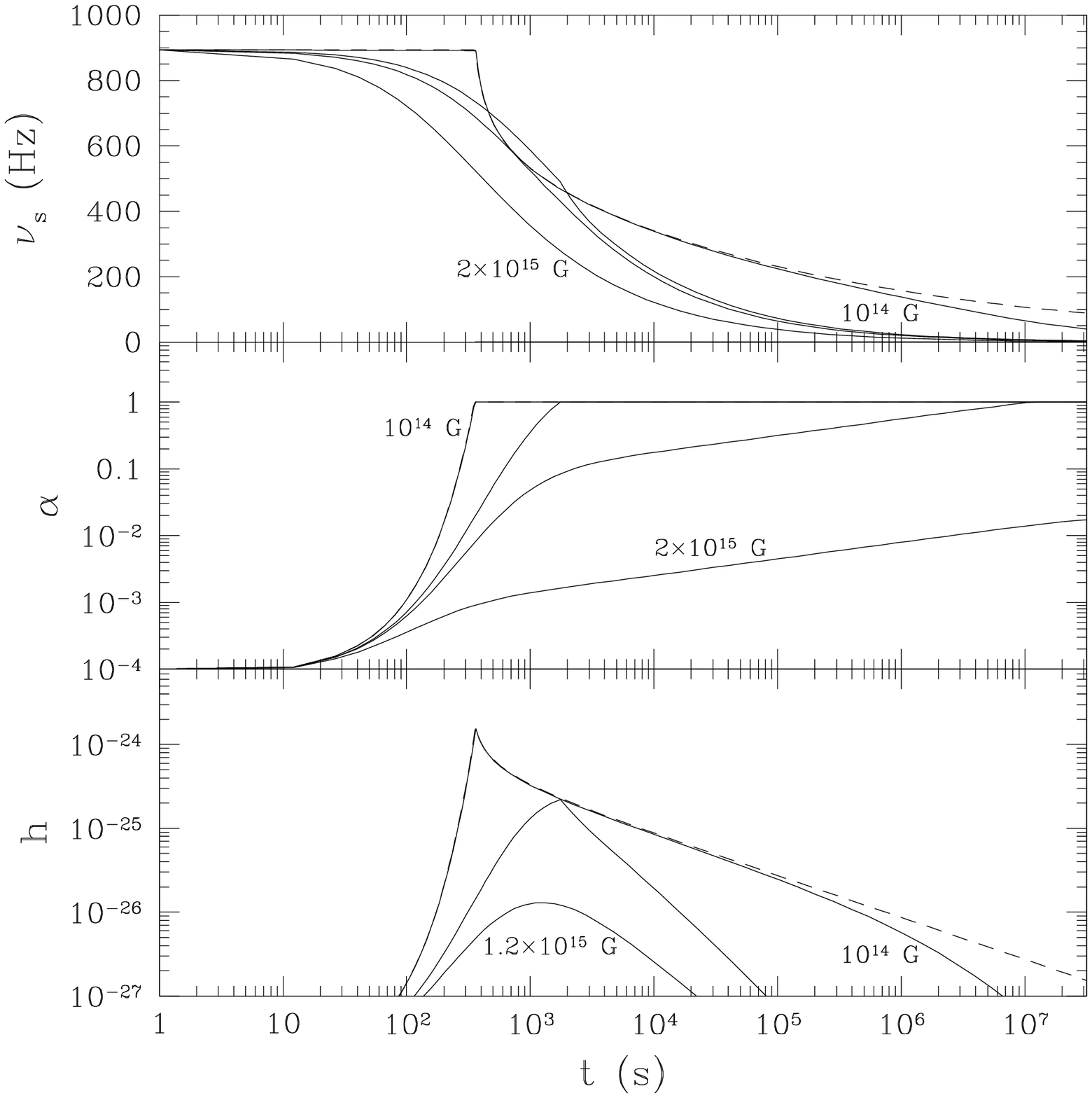}}
\caption{Spin frequency $\nus$, mode amplitude $\alpha$, and gravitational
wave amplitude $h$ as functions of time for a 1.4~$M_\odot$, 12.53~km
neutron star at $d=20$~Mpc with $\nusi = 890$~Hz, $\alphai = 10^{-4}$,
and $\kappa = 1$.  The dashed line is for the spindown caused by just
gravitational radiation ($B=0$).  The solid lines are for the
spindown that also includes magnetic braking with $B=10^{14}$~G,
$10^{15}$~G, $1.2\times 10^{15}$~G, and $2\times 10^{15}$~G from top to
bottom, respectively. Note that the line for $h$ with $B=2\times 10^{15}$~G is
not shown because it is below the level of interest.
\label{fig:time}}
\end{figure*}
\clearpage

\begin{figure*}
\epsfxsize=17cm
\epsfysize=17cm
\hbox{\hskip -0 truecm \epsffile{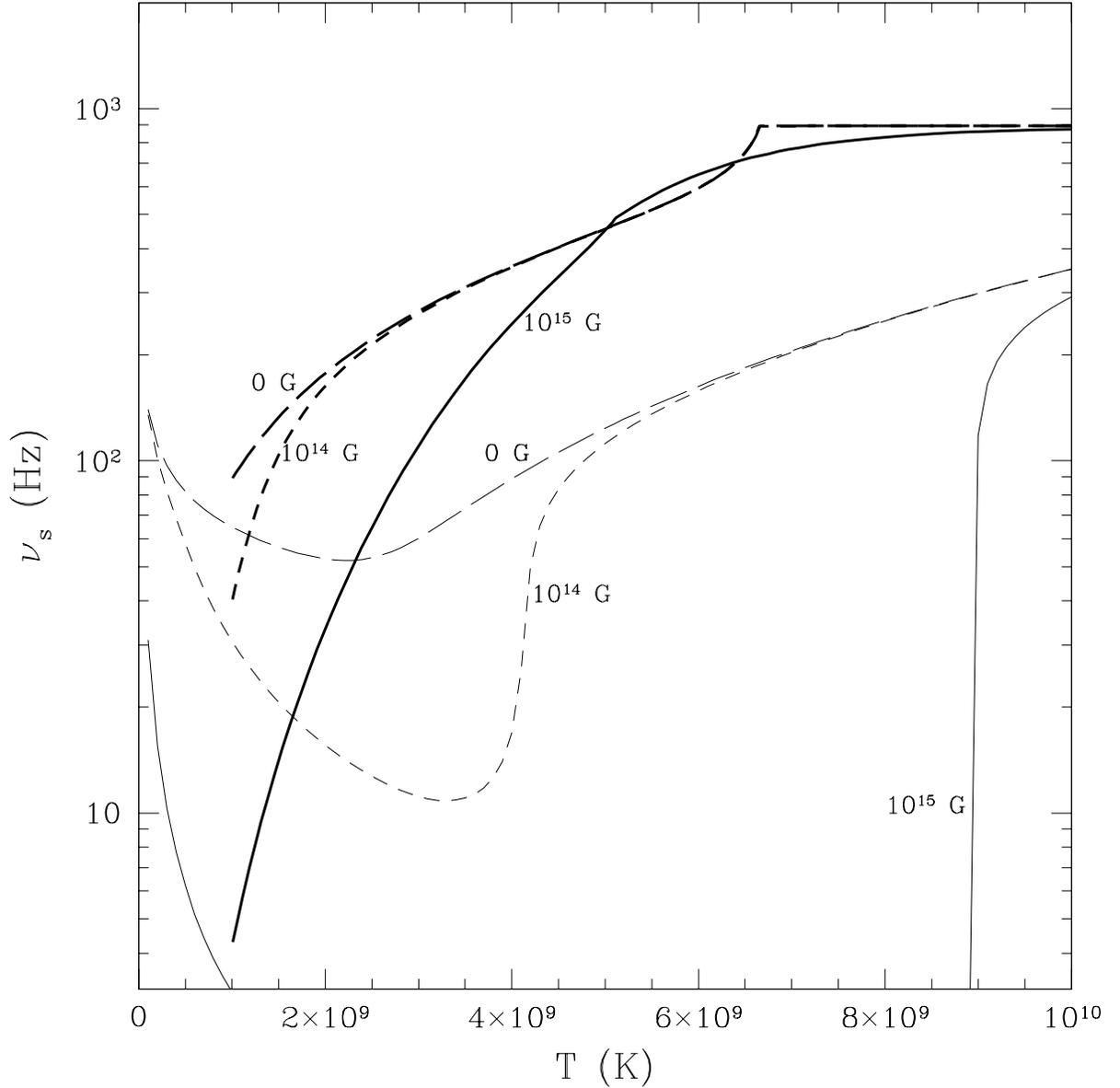}}
\caption{
Spin frequency $\nus$ as a function of the temperature T
for a 1.4~$M_\odot$, 12.53~km neutron star
with $\nusi = 890$~Hz, $\alphai = 10^{-4}$, and $\kappa = 1$.
The thick long dashed line is for the spindown caused by just
gravitational radiation ($B=0$). The thick short dashed line and
thick solid line are for the spindown that also includes magnetic braking
with $B=10^{14}$~G and $B=10^{15}$~G, respectively.
Also shown are the critical spin frequencies above which the r-modes
are unstable to growth.  These are denoted by thin lines corresponding
to the different magnetic field strengths.
\label{fig:temp}}
\end{figure*}
\clearpage

\begin{figure*}
\epsfxsize=17cm
\epsfysize=17cm
\hbox{\hskip -0 truecm \epsffile{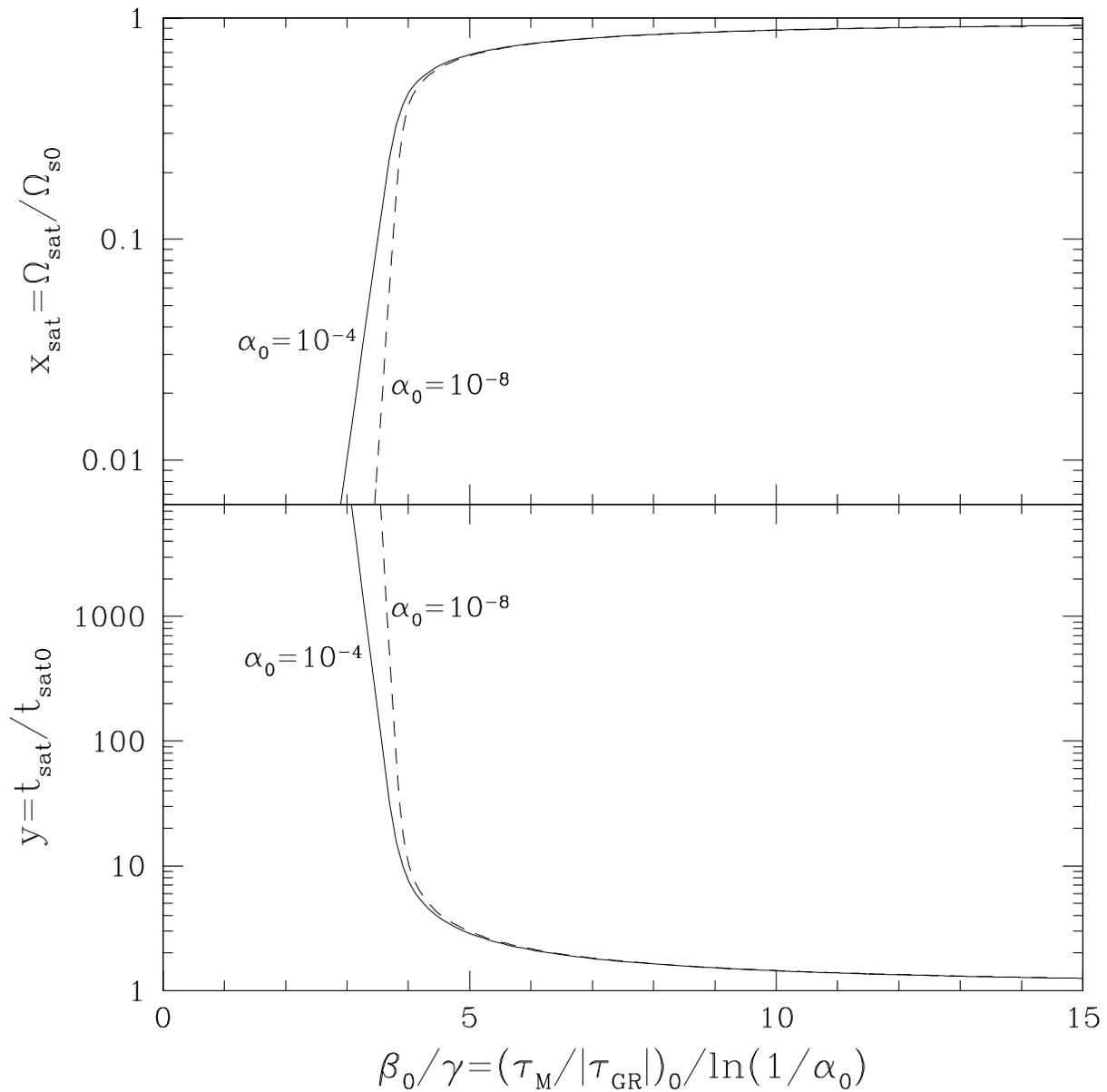}}
\caption{
$\xsat$ and $\ysat$ as functions of $\betai/\gamma$
for a 1.4~$M_\odot$, 12.53~km neutron star with $\nusi = 890$~Hz.
The solid line is for $\alphai = 10^{-4}$, and the dashed line is for $\alphai
= 10^{-8}$. See \S 2.2.
\label{fig:beta}}
\end{figure*}
\clearpage

\begin{figure*}
\epsfxsize=17cm
\epsfysize=17cm
\hbox{\hskip -0 truecm \epsffile{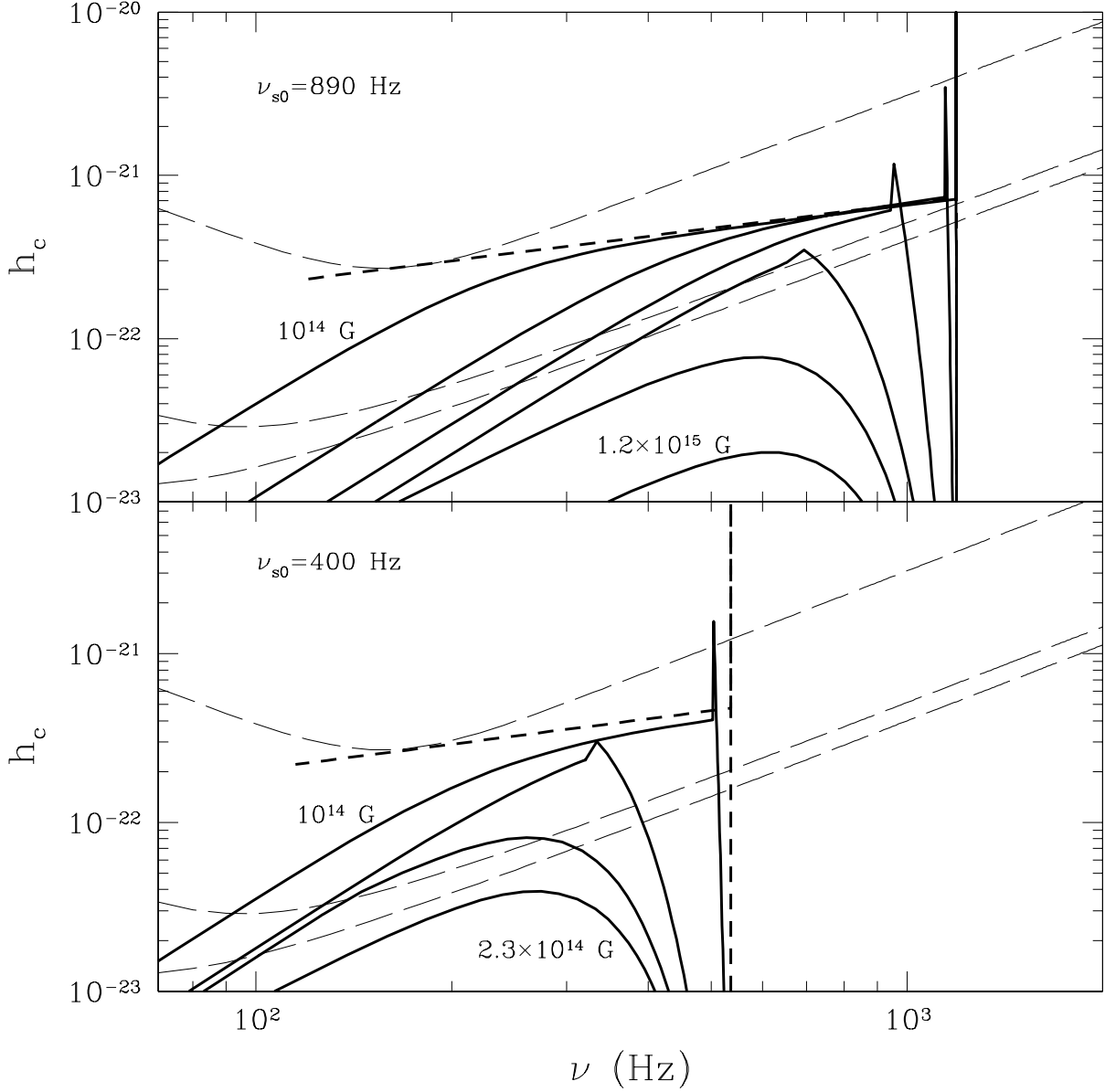}}
\caption{
Characteristic gravitational wave amplitude $h_{\rm c}=\nui\th$
as a function of frequency $\nui$ for a 1.4~$M_\odot$,
12.53~km neutron star at $d=20$~Mpc with $\nusi = 890$~Hz and
$\nusi = 400$~Hz,
$\alphai = 10^{-4}$, and $\kappa = 1$.  The short dashed line is for
the spindown caused by just gravitational radiation ($B=0$).
The solid lines are for the spindown that also includes magnetic braking.
For $\nusi = 890$~Hz, the magnetic fields are $B=10^{14}$~G,
$4\times 10^{14}$~G, $8\times 10^{14}$~G, $1\times 10^{15}$~G,
$1.1\times 10^{15}$~G, and $1.2\times 10^{15}$~G
from top to bottom, respectively.
For $\nusi = 400$~Hz, the magnetic fields are $B=10^{14}$~G,
$2\times 10^{14}$~G, $2.2\times 10^{14}$~G, and $2.3\times 10^{14}$~G
from top to bottom, respectively.
The long dashed lines are the noise amplitude
$h_{\rm rms}\equiv\sqrt{\nu S_h\lp\nu\rp}$ for LIGO.
\label{fig:hc}}
\end{figure*}

\end{document}